\documentclass[aps,prl,twocolumn,superscriptaddress,longbibliography]{revtex4-1}

\usepackage{graphicx}
\usepackage{bm}
\usepackage{hyperref}
\usepackage{todonotes}
\usepackage{color}
\usepackage{comment}
\usepackage{verbatim}
\usepackage{soul}

\begin{document}

\title{Doping Dependence of Collective Spin and Orbital Excitations in Spin 1 Quantum Antiferromagnet La$_{2-x}$Sr$_x$NiO$_4$ Observed by X-rays}

\author{G. Fabbris}
\email{gfabbris@bnl.gov}
\author{D. Meyers}
\affiliation{Department of Condensed Matter Physics and Materials Science, Brookhaven National Laboratory, Upton, New York 11973, USA}

\author{L. Xu}
\author{V. M. Katukuri}
\author{L. Hozoi}
\affiliation{Institute for Theoretical Solid State Physics, IFW Dresden, Helmholtzstra{\ss}e, 20, 01069 Dresden, Germany}

\author{X. Liu}
\affiliation{Beijing National Laboratory for Condensed Matter Physics and Institute of Physics, Chinese Academy of Sciences, Beijing 100190, China}
\affiliation{Collaborative Innovation Center of Quantum Matter, Beijing, China}

\author{Z.-Y. Chen}
\author{J. Okamoto}
\affiliation{National Synchrotron Radiation Research Center, Hsinchu 30076, Taiwan}

\author{T. Schmitt}
\affiliation{Research Department ``Synchotron Radiation and Nanotechnology'', Paul Scherrer Institut, CH-5232 Villigen PSI, Switzerland}

\author{A. Uldry}
\author{B. Delley}
\affiliation{Condensed Matter Theory Group, Paul Scherrer Institut, CH-5232 Villigen PSI, Switzerland}

\author{G. D. Gu}
\affiliation{Department of Condensed Matter Physics and Materials Science, Brookhaven National Laboratory, Upton, New York 11973, USA}

\author{D. Prabhakaran}
\author{A. T. Boothroyd}
\affiliation{Department of Physics, University of Oxford, Clarendon Laboratory, Oxford, OX1 3PU, UK}

\author{J. van den Brink}
\affiliation{Institute for Theoretical Solid State Physics, IFW Dresden, Helmholtzstra{\ss}e, 20, 01069 Dresden, Germany}

\author{D. J. Huang}
\affiliation{National Synchrotron Radiation Research Center, Hsinchu 30076, Taiwan}
\affiliation{Department of Physics, National Tsing Hua University, Hsinchu 30013, Taiwan}

\author{M. P. M. Dean}
\email{mdean@bnl.gov}
\affiliation{Department of Condensed Matter Physics and Materials Science, Brookhaven National Laboratory, Upton, New York 11973, USA}

\def\LNO{La$_2$NiO$_4$}
\def\LSNO{La$_{2-x}$Sr$_x$NiO$_4$}
\def\mathbi#1{\ensuremath{\textbf{\em #1}}}
\newcommand*{\hatH}{\hat{\mathcal{H}}}

\date{\today}

\begin{abstract}
We report the first empirical demonstration that resonant inelastic x-ray scattering (RIXS) is sensitive to \emph{collective} magnetic excitations in $S=1$ systems by probing the Ni $L_3$-edge of La$_{2-x}$Sr$_x$NiO$_4$ ($x = 0, 0.33, 0.45$). The magnetic excitation peak is asymmetric, indicating the presence of single and multi spin-flip excitations. As the hole doping level is increased, the zone boundary magnon energy is suppressed at a much larger rate than that in hole doped cuprates. Based on the analysis of the orbital and charge excitations observed by RIXS, we argue that this difference is related to the orbital character of the doped holes in these two families. This work establishes RIXS as a probe of fundamental magnetic interactions in nickelates opening the way towards studies of heterostructures and ultra-fast pump-probe experiments.
\end{abstract}


\maketitle

Spin and orbital degrees of freedom in transition metal oxides lie at the heart of their fascinating properties, motivating decades of effort to characterize their behavior \cite{Khomskii2014Book}. In the past few years, resonant inelastic x-ray scattering (RIXS) has emerged as an important tool for probing these spin and orbital states, complementary to inelastic neutron scattering (INS), photoemission and x-ray absorption \cite{Zaliznyak2015, deGrootBook, Ament2011, Braicovich2010, LeTacon2011, Dean2012, Schlappa2012, DeanLSCO2013, Lee2014, Dean2015, Kim2012, Kim2014excitonic, Dean2016ultrafast}. $L$-edge RIXS can even measure spin interactions in heterostructures \cite{Dean2012, Minola2012, Dantz2016}, and ultra-fast laser-induced transient states \cite{Dean2016ultrafast}. The vast majority of these successes have, however, focused on spin (or pseudo-spin) $S=1/2$ materials such as cuprates \cite{Braicovich2010, LeTacon2011, Dean2012, Schlappa2012, DeanLSCO2013, Lee2014, Dean2015} or iridates \cite{Kim2012, Kim2014excitonic, Dean2016ultrafast} and how their electronic interactions evolve with doping. These, however, represent a special case as only one $\Delta m_s=1$ spin transition is allowed on a single atomic site (i.e.\ $m_s=-1/2 \rightarrow m_s=1/2$), which directly matches the photon angular momentum. The ability of RIXS to address the electronic interactions in higher spin state compounds via measuring their collective magnetic excitations is unproven, as, contrary to INS, RIXS processes in $S>1/2$ systems can include other transitions such as $\Delta m_s=2$ \cite{deGrootBook}. La$_2$NiO$_4$ shares the same structural motifs as cuprates and iridates and also forms an antiferromagnetic Mott insulator in its ground state; however, its $3d^8$ configuration stabilizes an $S=1$ state \cite{Kuiper1998}. Whether RIXS can offer additional insights into exact nature of this state and how it evolves with doping remains largely unexplored. Indeed, the only available experimental work on $S \neq 1/2$ transition metal oxides focuses on $S = 1$ nickelate NiO and asserts that RIXS couples to \emph{local}  $\Delta m_s = 1$ and 2 spin flips, rather than \emph{collective} excitations \cite{Ghiringhelli2009, Chiuzbaian2005}, in line with influential early theoretical work that motivated the use of RIXS to access magnetic properties \cite{deGroot1998}.

\begin{figure*}
\includegraphics{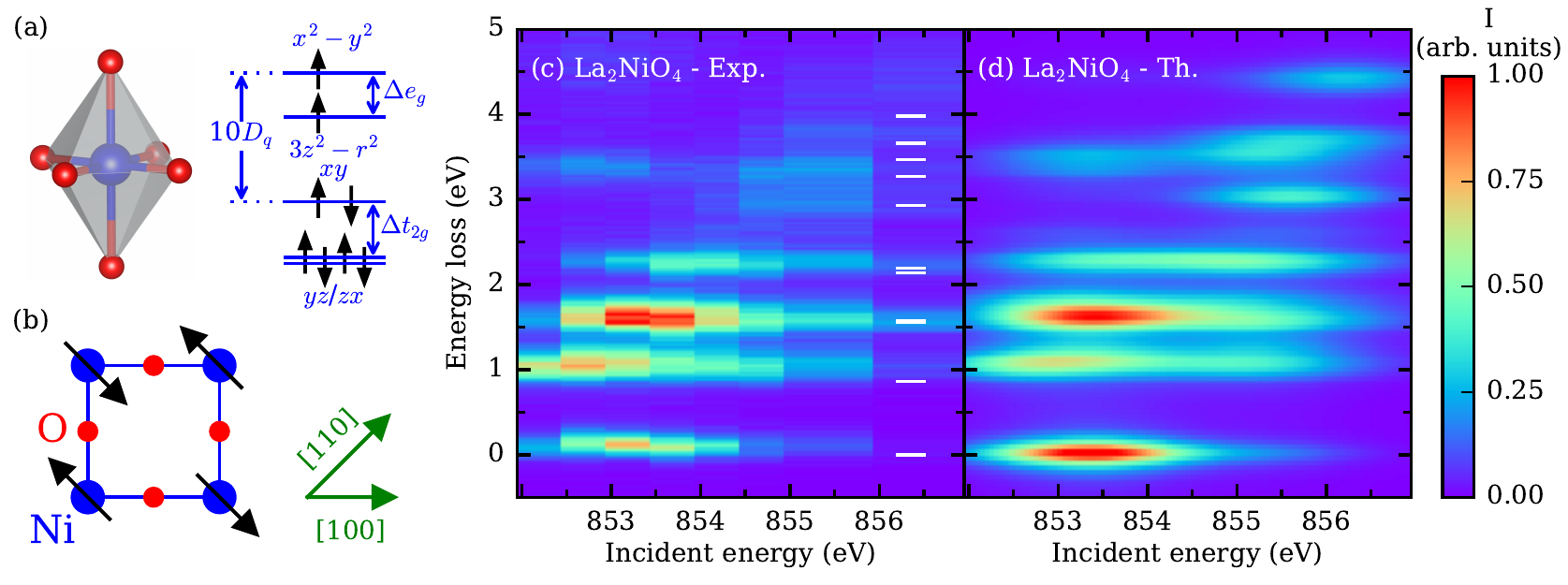}
\caption{(a) Depicts the tetragonally elongated NiO$_6$ octahedra present in \LSNO{}, which implies the energy level diagram plotted in blue. The electrons, in black, are found to populate these levels in a $3d^8$ $S=1$ configuration. (b) shows the known antiferromagnetic ordering of these spins \cite{Aeppli1988, Rodriguez-Carvajal1991}.  (c) \LNO{} Ni $L_3$-edge RIXS energy map collected at $Q_{\parallel} = (0.74\pi,0)$. White bars correspond to relative energies as computed by multireference configuration-interaction \cite{supplemental}. (d) Ni $L_3$ edge RIXS atomic multiplet calculation using parameters described in the text.}
\label{dd_maps}
\end{figure*}

In this Letter we present Ni $L_3$-edge RIXS measurements of the 2D antiferromagnet \LSNO{} (LSNO). Our central result is a direct  demonstration that RIXS can access \emph{collective} magnetic excitations in $S=1$ transition metal oxides, which is consistent with INS  \cite{Nakajima1993}, and which we exploit to examine the electronic evolution of the nickelates with doping. Furthermore, \emph{ab-initio} and atomic multiplet RIXS simulations closely reproduce the orbital excitations observed in the parent compound, confirming its localized $3d^8$ $S=1$ character, and providing a precise description of its crystal fields. Hole doping significantly reduces the zone boundary magnon energy ($\gtrsim 50\%$ at $x = 0.45$), consistent with INS work \cite{Nakajima1993, Yamada1991, Boothroyd2003, Boothroyd2003b, Bourges2003, Freeman2005, Woo2005}. Such reduction however is at odds with results from hole doped cuprates, for which the zone boundary magnetic energy scale is very weakly doping dependent  \cite{Vignolle2007,Lipscombe2007, DeanLSCO2013,LeTacon2013, Jia2014, Wakimoto2015}. We make use of RIXS sensitivity to orbital and charge excitations to infer that a larger $3d$ character of the doped holes in LSNO (when compared to cuprates) drives the magnetic energy scale reduction.

Single crystals of \LSNO{} ($x = 0, 0.33, 0.45$) were grown by the floating zone method \cite{Prabhakaran2002} and cleaved in vacuum immediately before measurements. Ni $L_3$-edge RIXS experiments were performed at low temperatures ($\sim20$~K) using the AGS-AGM \cite{Lai2014} and SAXES \cite{Ghiringhelli2006, Strocov2010} spectrometers \cite{supplemental}. Tetragonal notation with $a = b = 3.85$~\AA{} is used to describe in-plane wavevectors $Q_{\parallel}$   \cite{Hucker2004}. The combined energy resolution for the $x = 0$ data is $\sim 150$~meV full width half maximum, while $\sim 100$~meV was achieved for $x = 0.33$ and $0.45$ \footnote{The AGS-AGM beamline was better optimized during the measurements on the $x = 0.33$ and $0.45$ samples, leading to a better energy resolution in the RIXS data}. The zero energy loss position was calibrated for every spectrum by measuring a carbon tape. 

We first address the orbital configuration of LSNO parent compound. La$_2$NiO$_4$ is isostructural to the high-$T_c$ superconductor La$_2$CuO$_4$, with tetragonally distorted NiO$_6$ octahedra as shown in Fig. \ref{dd_maps}(a) \cite{Hucker2004}. A RIXS map plotting the incident energy dependence of the orbital excitations of La$_2$NiO$_4$ is displayed in Fig.~\ref{dd_maps}(c). The presence of well-defined constant energy loss $dd$ excitations is strong evidence for the localized character of the Ni $3d$ states. To analyze these results, we first computed the orbital excitation energies from first principles using multireference configuration interaction calculations \cite{Helgaker2000}. The energies [white bars in Fig.~\ref{dd_maps}(c)] match the experimental values within the expected accuracy of $\sim 10-15$\% \cite{Huang2011} and justified modeling the data based on a high-spin $S=1$ $3d^8$ ground-state, similar to previous analysis of x-ray absorption spectroscopy \cite{vanderLaan1988,vanderLaan1988,Kuiper1998}. We then performed semi-phenomenological atomic calculations to extract crystal field values based on maximizing the agreement between the calculations and the data \cite{deGroot2005, deGrootBook, Stavitski2010, Uldry2012,  Fabbris2016, supplemental}. The final result, plotted in Fig.~\ref{dd_maps}(d), captures the observed peak intensity and resonant behavior. The extracted crystal field splittings, as defined in Fig.~\ref{dd_maps}(a), are $10D_q = 1.6 \pm 0.1$~eV, $\Delta e_g = 0.75 \pm 0.05$~eV, and $\Delta t_{2g} = 0.1 \pm 0.05$~eV \footnote{Best agreement between data and atomic calculations was obtained with reduced Slater-Condon parameters. $F_{dd} = 65\%$, $F_{pd} = 65\%$, and  $G_{pd} = 85\%$. We note that the reported crystal fields correspond to the effective energies, which includes hybridization effects. Pre-hybridization energies, as those obtained from a charge transfer model, are expected to be different.}. The rather large $\Delta e_g$ is overcome by electron-electron interaction driving the $S=1$ state \cite{VanderLaan1986}. Disagreement between experimental and calculated spectra primarily occurs at higher energy loss, which is likely a consequence of an intensity renormalization due to charge transfer excitations that is not included in the present model \cite{Ghiringhelli2005}, but that occurs at $\approx 7.5$~eV energy loss \cite{supplemental}.

\begin{figure}
\includegraphics{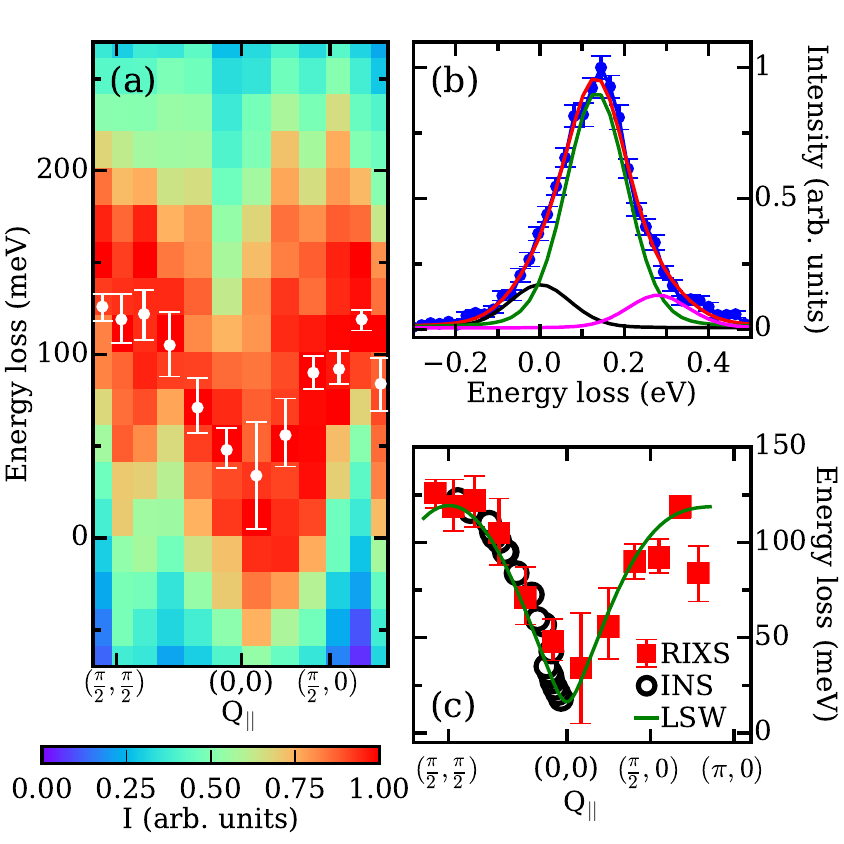}
\caption{(a) \LNO{} low energy excitations $Q_{\parallel}$ dependence map collected at $E_i$ = 853.2 eV. White circles correspond to the fitted magnon energies. (b) Fitting example at $Q_{\parallel} = (0.48\pi, 0.48\pi)$, black, green and magneta lines account for elastic, single magnon and multi-magnon excitations similar to previous RIXS analysis \cite{Dean2015}. (c) \LNO{} magnon dispersion (red squares) compared to inelastic neutron scattering results (black circles) and spin wave theory fits (green line) \cite{Nakajima1993, RIXS_INS_comp}. The error bars shown in panels (a) and (c) correspond to 95\% confidence intervals obtained from the least square fitting algorithm.}
\label{x0}
\end{figure}

Figure~\ref{x0} examines the momentum dependence of the low-energy excitations in La$_2$NiO$_4$ showing a peak that, based on its energy scale and dispersion, is assigned to a spin wave or magnon excitation. As shown in Fig.~\ref{x0}(b), and discussed in detail later, the spectral lineshape is most naturally fit by a three component model with the width of each peak set to the energy resolution. The dispersion of the strongest peak is in excellent agreement with INS confirming its assignment as a magnon [Fig.~\ref{x0}(c)] \cite{Nakajima1993}. This is the first RIXS measurement of \emph{dispersive} magnetic excitations in a $S \neq 1/2$ transition metal oxide and is very significant as it enables the use of RIXS to investigate magnetic interactions in systems for which INS experiments remain challenging. Such a result seems very natural in view of the extensive observations of magnons in $S=1/2$ local moment materials such as cuprates and iridates \cite{Ament2011, Braicovich2010, LeTacon2011, Dean2012, Schlappa2012, DeanLSCO2013, Lee2014, Dean2015, Kim2012, Kim2014excitonic, Dean2016ultrafast}. It does, however, contradict the existing experimental literature regarding how RIXS couples to magnetic excitations in $S=1$ materials \cite{Ghiringhelli2009}. In a localized Ni $3d^8$ $S=1$  triplet [Fig. \ref{dd_maps}(a)], $\Delta m_s=1,2$ transitions can in principle be obtained in a local perspective \cite{deGroot1998}, breaking the one-to-one correspondence between the allowed on-site spin transitions and magnons, and complicating the issue of whether RIXS accesses collective excitations. Indeed, previous studies of NiO have asserted that Ni $L_3$ RIXS is sensitive to \emph{local} spin flips, rather than \emph{collective} magnons \cite{Ghiringhelli2009}. 

Given that single magnon excitations are present, one would expect multi-magnon excitations also to occur, with higher energy scales and much weaker \mathbi{Q}-dependence \cite{Haverkort2010}. A combination of three pseudo-Voigt energy resolution functions, corresponding to elastic, magnon and multi-magnon peaks, provided the simplest means to adequately fit the data particularly in view of similar approaches applied to the cuprates  \cite{Braicovich2010,LeTacon2011,Dean2012,DeanLSCO2013}. The energy resolution of the present data is insufficient to extract the multi-magnon spectral lineshape precisely, thus the spin process leading to the multi-magnon intensity cannot be unambiguously determined since both $\Delta m_s = 0$ and $\geq 2$ are possible. Nevertheless, the data are best fit by fixing the energy of the multi-magnon peak to twice the zone boundary magnon energy (252~meV), as theoretically suggested \cite{Haverkort2010}, indicating that it corresponds to $\Delta m_s = 2$ magnons. The results are shown in Fig.~\ref{x0}(c). We found that the zone boundary magnetic excitations energies at $(\pi,0)$ and $(\pi/2,\pi/2)$ were, within error, consistent with one another, indicating a small next-nearest-neighbor magnetic exchange ($\lesssim$10~meV) or $\lesssim$8\% of the overall energy scale. Such exchange is substantially smaller than in cuprates (20\%) \cite{Coldea2001, Dean2012} and iridates (60\%) \cite{Kim2012}, but larger than observed in cobaltates (0.6\%) \cite{Babkevich2010}, suggesting a reduced influence of long range magnetic coupling in lighter transition metal oxides. Therefore, following Nakajima \emph{et al.} \cite{Nakajima1993}, we used a magnetic Hamiltonian containing first neighbor exchange, $J_1$, and $c$-axis anisotropy fixed at $J_c = 0.52$~meV

\begin{equation}
    \hatH = J_1 \sum_{\left \langle i,j \right \rangle} \vec{S}_i \cdot \vec{S}_j + J_c \sum_{i} (S_i^c )^2 .
\end{equation}

Fitting of $J_1$ within spin wave theory yields $27\pm 1$~meV, similar to the previous value of $28.7\pm0.7$~meV \cite{Nakajima1993}. The close agreement observed between RIXS, INS and linear spin wave modeling is further proof of the ability of RIXS to probe \emph{collective} magnon excitations in systems with localized $3d$ states, as predicted based on effective-operator theory calculations for a similar $S=1$ $d^8$ model \cite{Haverkort2010}. As a further check, we computed the strength of the exchange interaction coupling neighboring Ni $3d$ orbitals via the in-plane O $2p$ orbitals using difference-dedicated configuration interaction (DDCI) calculations \cite{Miralles1993} (see Supplemental Material \cite{supplemental}). We find $J_1 = 22.3$~meV, 17\% lower than experiment. Higher values are expected by additionally including the apical O 2$p$ orbitals in the DDCI treatment.

\begin{figure}[t]
\includegraphics[width = 7.5 cm]{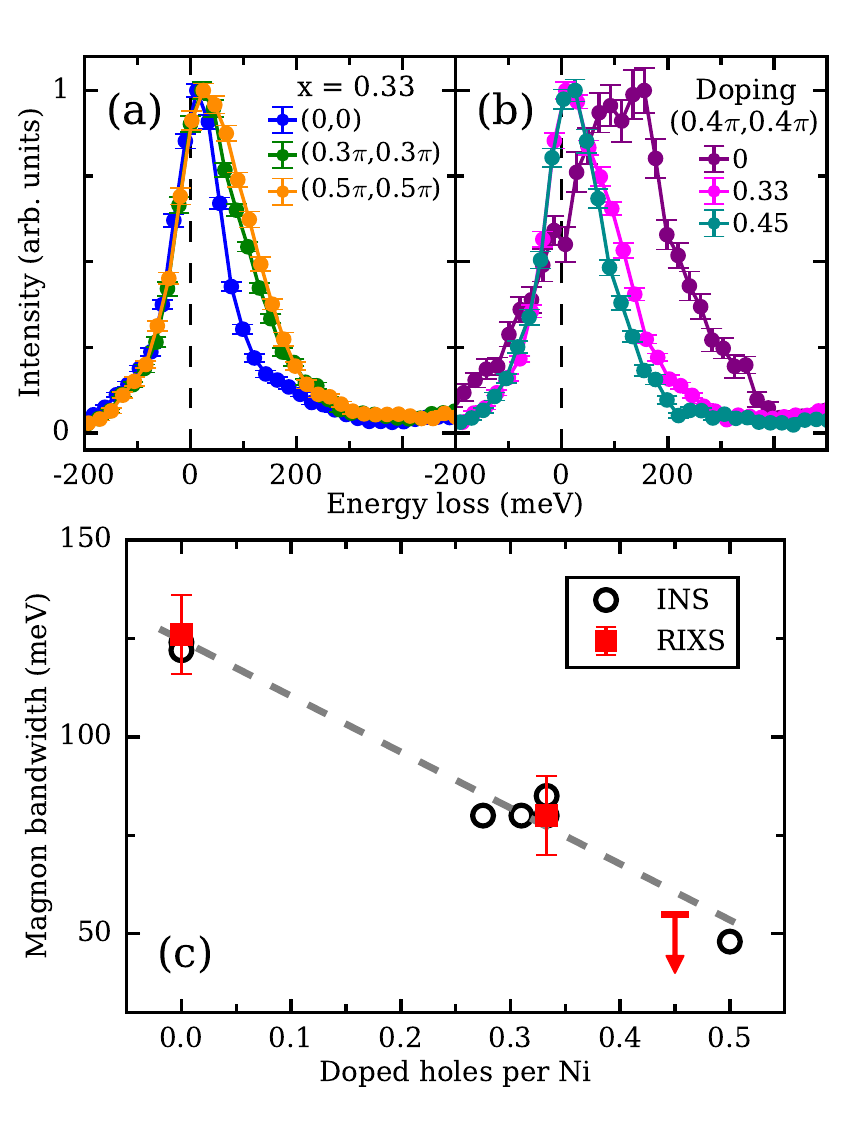}
\caption{Doping dependence of the magnetic excitations. (a) Even though the magnon bandwidth is significantly smaller at $x = 0.33$, a distinct dispersion is observed. (b) Doping dependence of the low energy excitations at $Q_{||}$ = $(0.39\pi,0.39\pi)$. No clear magnetic excitation is observed for $x = 0.45$. (c) Doping dependence of magnon bandwidth compared to results from INS \cite{Nakajima1993, Yamada1991, Boothroyd2003, Boothroyd2003b, Bourges2003, Freeman2005, Woo2005}.}
\label{magnon_doping}
\end{figure}

We now examine the doping dependence of magnetic excitations. Figure \ref{magnon_doping}(a) plots the data for $x=0.33$. Despite the significant bandwidth reduction, a dispersing feature is observed, consistent with a magnon excitation. Using the same approach as for $x = 0$, a maximum magnon energy of $80 \pm 10$~meV is retrieved, consistent with INS \cite{collective, Boothroyd2003b}. We further plot the doping dependence of the peak at $Q_{\parallel} = (0.4\pi,0.4\pi)$ in Fig.~\ref{magnon_doping}(b) showing a substantial softening with doping. No clear magnetic excitation was observed in the $x = 0.45$ sample, indicating that any signal lies below $\sim55$~meV. Figure \ref{magnon_doping}(c) plots the energy scale of the magnetic peak as a function of doping showing a softening of $\gtrsim 50\%$ at $x=0.45$. This softening is substantially larger than that in doped cuprates, in which the magnetic bandwidth decreases very slowly with hole doping \cite{Dean2015, Lipscombe2007,Lipscombe2009, Braicovich2010, LeTacon2011, DeanLSCO2013, DeanBSCCO2013, DeanLBCO2013, Lee2014, Minola2015, Wakimoto2015}. Furthermore, this points a non-trivial evolution of the nickelate electronic structure beyond that of the single band nearest neighbor Hubbard model, as within this model cuprates and nickelates would be expected to be rather similar.

\begin{figure}[t]
\includegraphics[width = 7.5 cm]{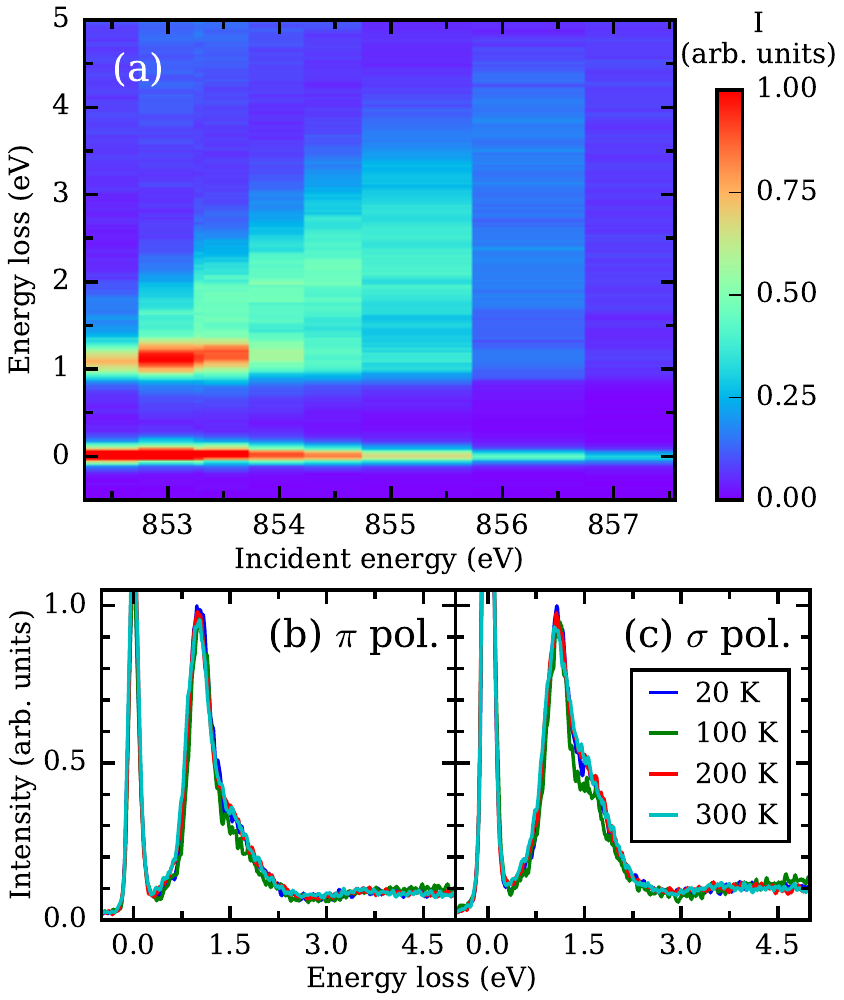}
\caption{Ni L$_3$ edge RIXS map for \LSNO{} $x = 1/3$ at $Q_{\parallel} = (0.74\pi,0)$. The localized $dd$-excitations observed for the $x=0$ parent compound [Fig.~\ref{dd_maps}(c)] are strongly suppressed with doping alongside x-ray fluorescence appearing as a broad diagonal line of intensity.(b)\&(c) RIXS spectra at 853.5 eV incident x-ray energy and 15$^{\circ}$ incident angle. In this geometry \cite{supplemental}, $\pi$ and $\sigma$ polarization primarily places the x-ray electric field along $a$ and $c$, respectively.}
\label{doped_dd_map}
\end{figure}

Further insight into the electronic state of \LSNO{} can be obtained by examining its charge and orbital excitations, as plotted in Fig.~\ref{doped_dd_map}(a). A dramatic suppression of localized $dd$ excitations is seen with respect to Fig.~ \ref{dd_maps}(c). Only a single Raman peak is observed at $\sim$1~eV together with a broad diagonal feature coming from x-ray fluorescence, indicating that Sr substitution significantly modifies the Ni $3d$ orbitals. The ground state of LSNO $x = 0.33$ can be conceptualized as the mixture $\alpha \vert 3d^7 \rangle + \beta \vert 3d^8 \rangle + \gamma \vert 3d^8 \underline{L} \rangle$, with the later being a Ni $3d$ - O $2p$ ligand hole. We performed multiplet RIXS calculations for an appropriate mixture of Ni $d^7$ and $d^8$, which we find do not reproduce the measured spectra \cite{supplemental}. Instead, the strong presence of fluorescence closely resembles the signal observed in NdNiO$_3$, which is also incompatible with single site atomic multiplet calculations, but consistent with $\vert 3d^8 \underline{L} \rangle$ states stabilized by the negative charge transfer energy \cite{Bisogni2016}. The similar phenomenology here implies that $\vert 3d^8 \underline{L} \rangle$ is also the dominant state in LSNO $x = 0.33$.  Finally, we further studied the temperature dependence of the excitations, finding that a similar spectra persist despite the charge and spin stripe phase transitions at 240~K and 190~K, respectively \cite{Lee1997} and a large change in optical conductivity \cite{Katsufuji1996}. This is in contrast to earlier reports using Ni K-edge RIXS \cite{Simonelli2010}, and shows that the high temperature phase of LSNO retains a very similar local orbital configuration likely due to persistent short-range dynamic stripe correlations \cite{Abeykoon2013, Anissimova2014,Zhong2016}.

It is notable that cuprates, which also have a negative charge transfer energy, show a far smaller doping dependence of $dd$ excitations than that seen here in LSNO \cite{DeanLSCO2013, Meyers2017, supplemental}. This can be rationalized by noting that the ligand hole wavefunction corresponds to a mixture of the $3d$ and $2p$ orbitals. In both nickelates and cuprates this mixture is believed to be dominated by the $2p$ character \cite{Chen1991,Kuiper1995,Pellegrin1996}. However, hole doping LSNO largely disrupts the $3d$ atomic multiplet structure, which suggest that its ligand hole state has a larger $3d$ character than in cuprates. In fact, the large linear dicroism on the orbital excitations of $\mathrm{La_2NiO_4}$ is dramatically suppressed at $x = 0.33$ [See Fig. \ref{doped_dd_map} (b)\&(c) and Supplemental Material \cite{supplemental}], suggesting that the $\vert 3d^8 \underline{L} \rangle$ state has substantial contributions of both $3z^2-r^2$ and $x^2-y^2$ orbitals, a scenario that is further corroborated by ARPES results in highly doped samples \cite{Uchida2011b}. We therefore propose that the stronger magnon softening in nickelates, compared to cuprates, relates to larger Ni $3d$ character of the doped holes, with a possible further role for polaron formation in attenuating the strength of magnetic exchange.

In conclusion, we show that Ni $L$-edge RIXS is sensitive to \emph{collective} magnetic excitations. This is a key observation since it places RIXS in a prime position in the study of magnetic exchange interactions in systems and/or experimental setups that are incompatible with inelastic neutron scattering, such as thin film heterostructures and at ultra-fast timescales. Furthermore, we observe a significant suppression of the magnetic energy scale upon hole doping, an intriguing behavior since the magnon energy is weakly doping dependent in cuprates \cite{Dean2015,Lipscombe2007,Braicovich2010,LeTacon2011,DeanLSCO2013,DeanBSCCO2013,DeanLBCO2013,Minola2015,Wakimoto2015}. Analysis of RIXS orbital and charge excitations indicate that this behavior derive from a larger degree of $3d$ character in the doped holes wavefunctions of nickelates. RIXS has experienced a fast paced advance on experimental energy resolution and instrumentation over the last decade \cite{Ament2011,Ghiringhelli2006,Lai2014,Dvorak2016}, and the demonstration of ultra-fast RIXS \cite{Dean2016ultrafast}. Together with advances in theoretical modeling, such capabilities will likely establish RIXS as a prime tool for condensed matter research.

\begin{acknowledgments}
We thank John M. Tranquada and Valentina Bisogni for valuable discussions. This material is based upon work supported by the U.S.\ Department of Energy, Office of Basic Energy Sciences, Early Career Award Program under Award Number 1047478. Work at Brookhaven National Laboratory was supported by the U.S. Department of Energy, Office of Science, Office of Basic Energy Sciences, under Contract No. DE-SC00112704. X.L. acknowledges financial support from MOST (No. 2015CB921302) and CAS (Grant No: XDB07020200) of China. Work at the Paul Scherrer Institut by T.S.\ has been funded by the Swiss National Science Foundation through the Sinergia network Mott Physics Beyond the Heisenberg (MPBH) model. Experiments were performed at BL05A1 - Inelastic Scattering at National Synchrotron Radiation Research Center, Taiwan \cite{Lai2014} and using the SAXES spectrometer \cite{Ghiringhelli2006} at the ADRESS beam line \cite{Strocov2010} of the Swiss Light Source at the Paul Scherrer Institut, Switzerland.
\end{acknowledgments}

\bibliography{refs}

\begin{thebibliography}{75}
\bibitem [{RIX()}]{RIXS_INS_comp}%
  \BibitemOpen
  \href@noop {} {}\bibinfo {note} {This comparison assumes that the dispersion
  measured with neutrons around $(\pi,\pi)$ is the same as that measured by
  RIXS around $(0,0)$ as expected in an antiferromagnet.}\BibitemShut {Stop}%
\end{thebibliography}

\end{document}